%% file: main.tex
\documentclass{llncs}
\usepackage{cite}
\usepackage{amsmath,amssymb,amsfonts}
\usepackage{algorithmic}
\usepackage{graphicx}
\usepackage{textcomp}
\usepackage{xcolor}
\usepackage{url}
\usepackage{gensymb}
\usepackage{siunitx}
\usepackage{censor}
\usepackage{lipsum}
\usepackage{balance}
\usepackage{float}
\usepackage{subcaption}
\usepackage{array}

\usepackage{tikz}

\usetikzlibrary{shapes.geometric, arrows.meta, positioning, fit}

\tikzset{
  personarrow/.style={->, >={Latex[length=2mm,width=2mm]}, line width=0.5mm, draw=ibmblue},
  purplearrow/.style={->, >={Latex[length=2mm,width=2mm]}, line width=0.5mm, draw=ibmpurple},
  pinkarrow/.style={->, >={Latex[length=2mm,width=2mm]}, line width=0.5mm, draw=ibmpink},
  orangearrow/.style={->, >={Latex[length=2mm,width=2mm]}, line width=0.5mm, draw=ibmorange},
  orangeline/.style={-, line width=0.5mm, draw=ibmorange},
  yellowarrow/.style={->, >={Latex[length=2mm,width=2mm]}, line width=0.5mm, draw=ibmyellow},
  grayarrow/.style={->, >={Latex[length=2mm,width=2mm]}, line width=0.5mm, draw=mygray},
  timecaption/.style={inner sep=1mm, draw},
  passenger-blue/.style={inner sep=1mm, node contents={\includegraphics[width=3mm]{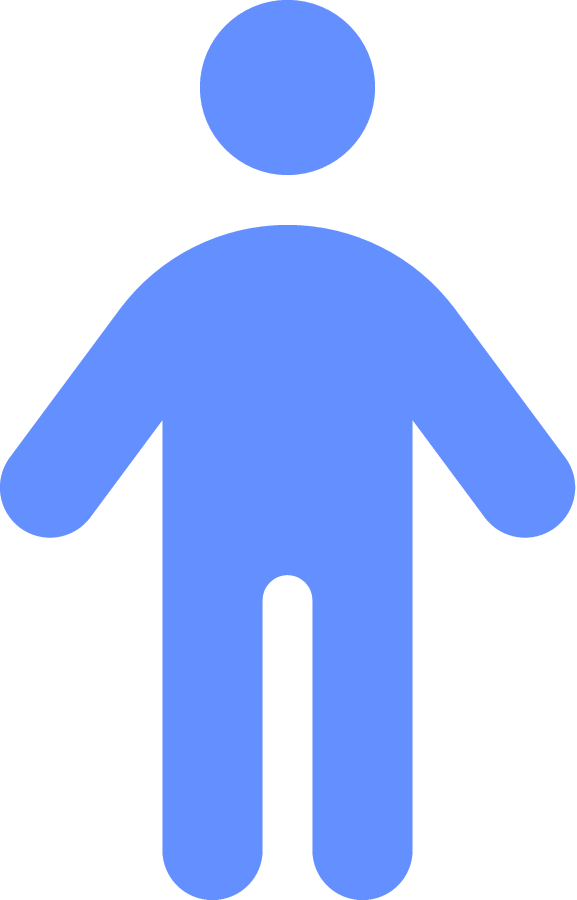}}},
  house-blue/.style={inner sep=1mm, node contents={\includegraphics[width=4mm]{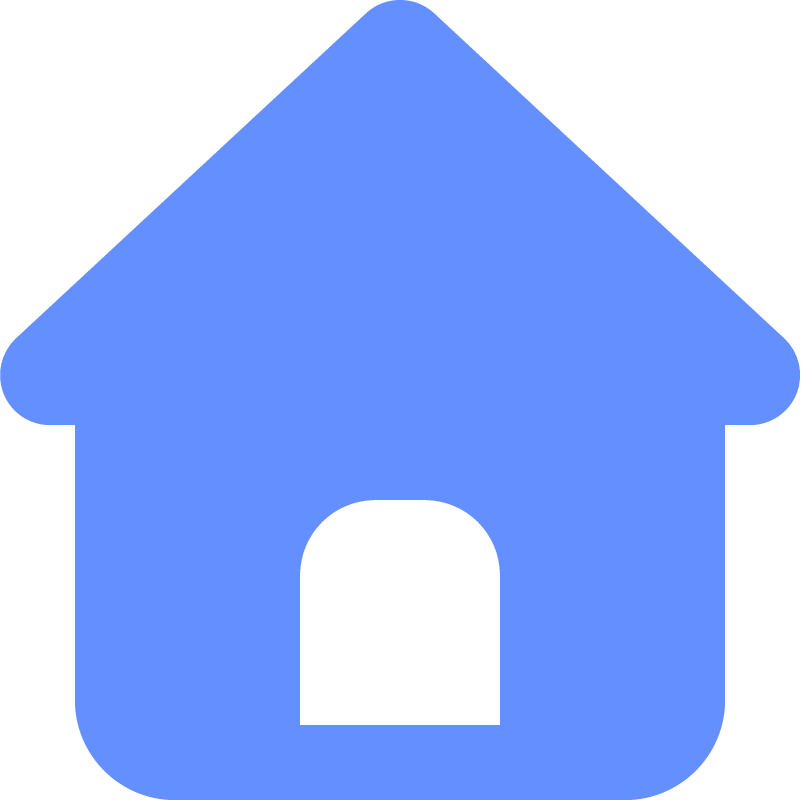}}},
  van-shuttle-yellow/.style={inner sep=1mm, node contents={\includegraphics[width=4mm]{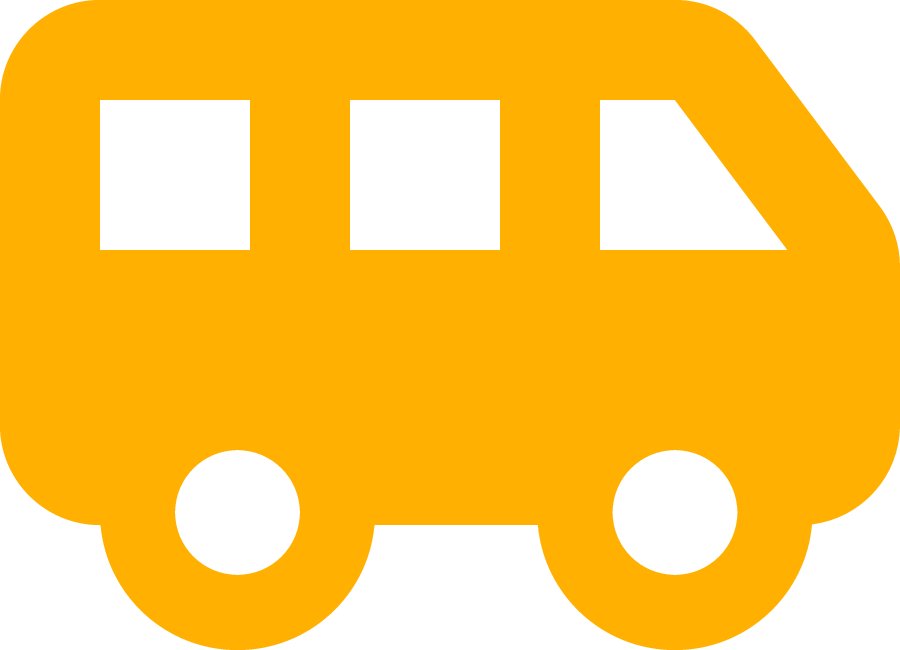}}},
  van-shuttle-pink/.style={inner sep=1mm, node contents={\includegraphics[width=4mm]{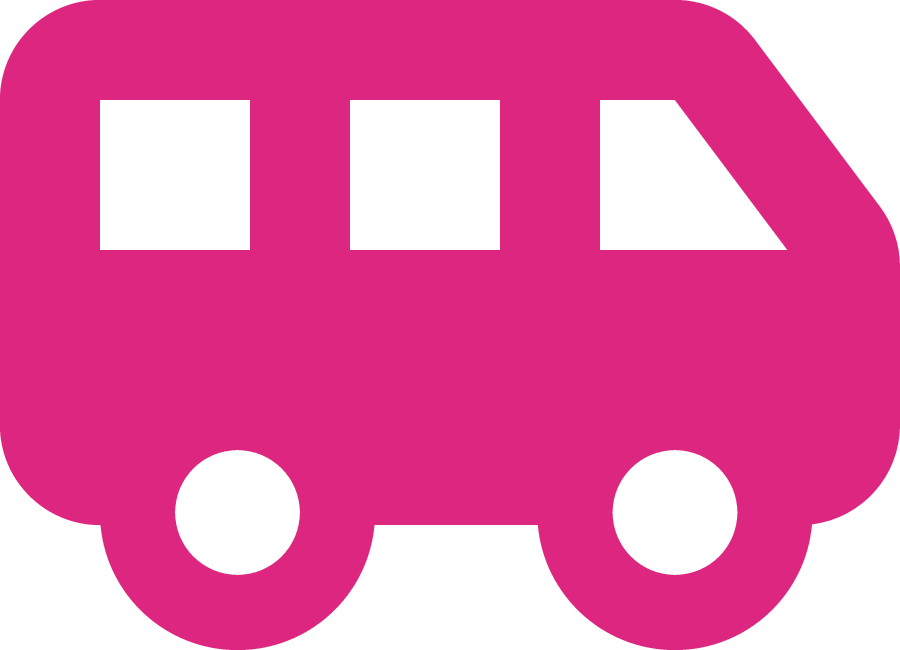}}},
  van-shuttle-orange/.style={inner sep=1mm, node contents={\includegraphics[width=4mm]{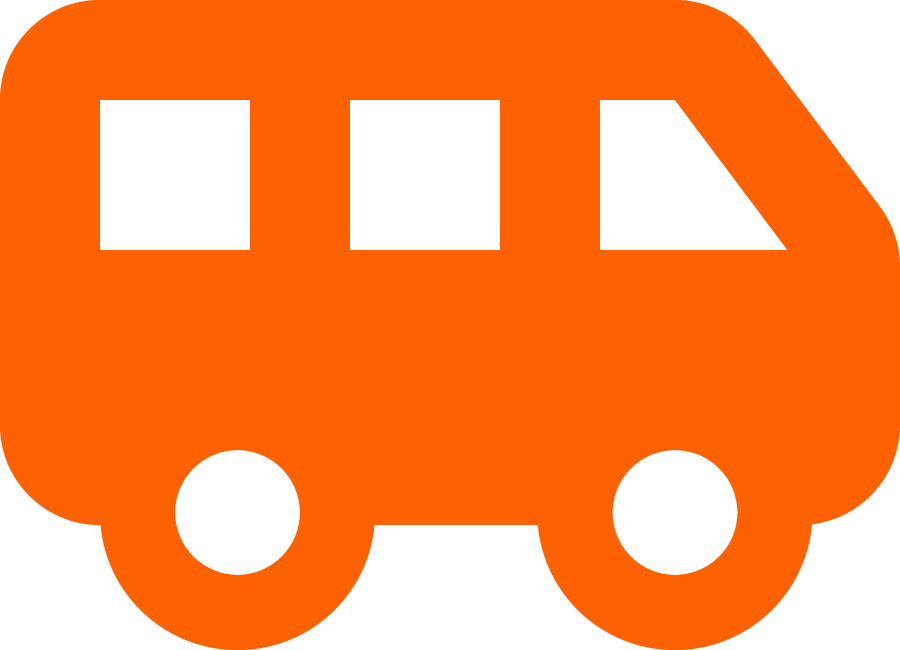}}},
  van-shuttle-purple/.style={inner sep=1mm, node contents={\includegraphics[width=4mm]{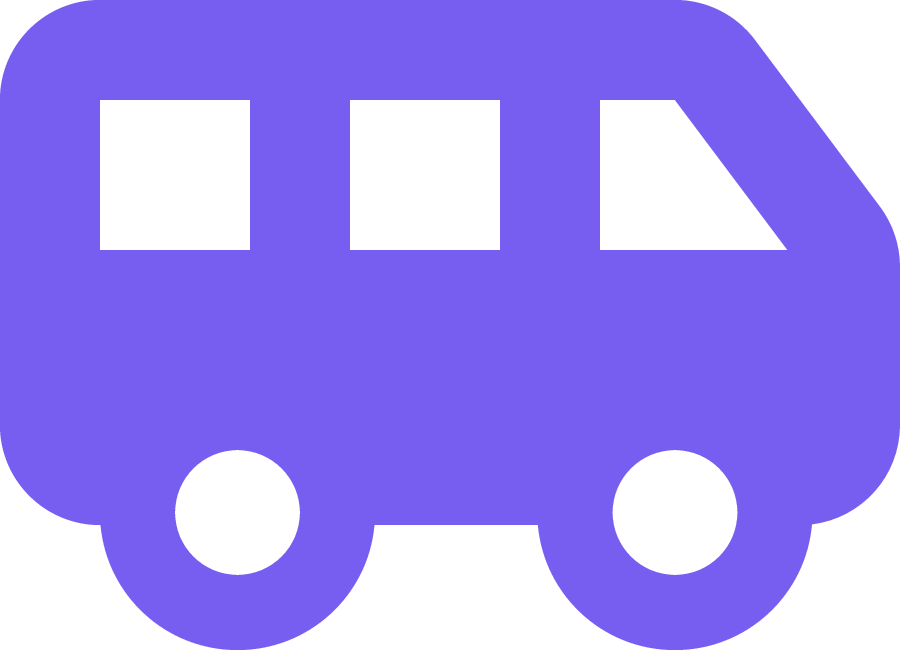}}},
  van-shuttle-gray/.style={inner sep=1mm, node contents={\includegraphics[width=4mm]{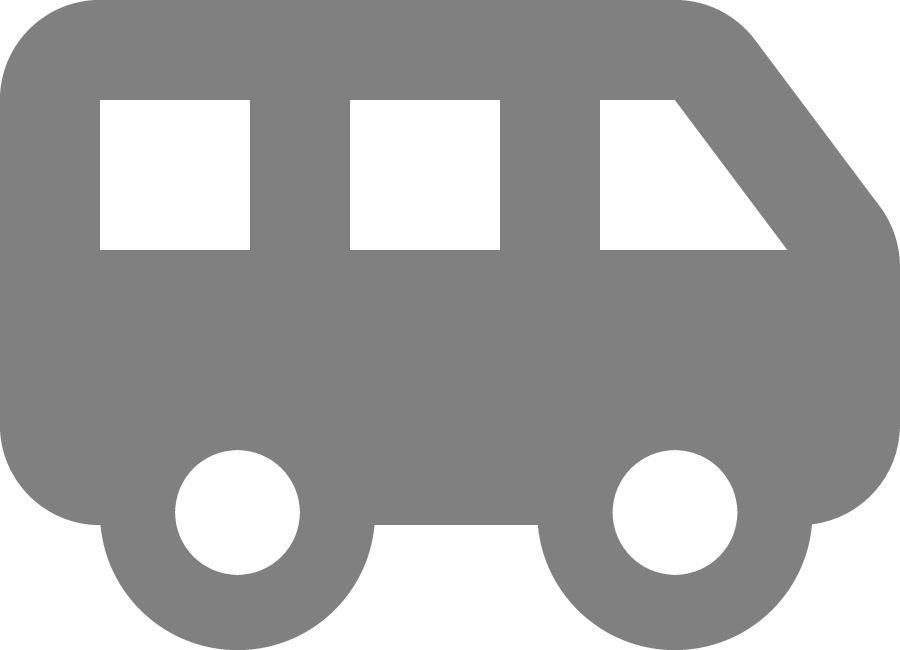}}},
}

\definecolor{ibmblue}{HTML}{648FFF}
\definecolor{ibmpurple}{HTML}{785EF0}
\definecolor{ibmpink}{HTML}{DC267F}
\definecolor{ibmorange}{HTML}{FE6100}
\definecolor{ibmyellow}{HTML}{FFB000}
\definecolor{mygray}{HTML}{808080}

\newcolumntype{?}[1]{!{\vrule width #1}}

\usepackage[bookmarks]{hyperref}

\newcommand{\rev}[1]{{\color{.} #1}}

\title{A Swarm Approach to Public Transit Using On-demand Routing in a Slime-Mold-Inspired Framework}
\titlerunning{Swarm Transit}

\author{Lindsay Burke\inst{1}\orcidID{0009-0005-0444-0585}
\and
Maxfield Comstock\inst{2}\orcidID{0000-0002-8765-9830}
\and
Jason M.\ Graham\inst{3}\orcidID{0000-0003-0047-7178}
\and
R.~F.~Malenda\inst{4}\orcidID{0000-0003-0434-3150}
\and
Simon Garnier\inst{2}\orcidID{0000-0002-3886-3974}
\and
Petras Swissler\inst{4}\orcidID{0000-0002-8528-4449}
}

\authorrunning{Burke et al.}

\institute{
Department of Computer Science, New Jersey Institute of Technology 
\and Federated Department of Biological Sciences, New Jersey Institute of Technology
\and Department of Mathematics, University of Scranton
\and Department of Mechanical and Industrial Engineering, New Jersey Institute of Technology
}

\date{March 2026}

\begin{document}
\maketitle

\begin{abstract}
Demand-responsive transit (DRT) is a flexible alternative to traditional, fixed-route mass-transit networks. Although DRT can function well in low-density communities, high operating costs and low reliability are common issues. We propose that these issues can be mitigated by moving from a centralized, manually-scheduled scheme to a distributed system capable of dynamically routing multiple vehicles using a slime-mold-inspired routing algorithm to maximize network effectiveness.
In this paper, we present simulated results for \rev{swarm-driven routing on a} transit network in urban, suburban, and semi-rural scenarios, using map networks pulled from OpenStreetMap. We show that our approach increases passenger delivery rates relative to a fixed-\rev{route} approach by \rev{\qty{56}{\percent}}, \rev{\qty{78}{\percent}}, and \rev{\qty{128}{\percent}}, respectively, and results in over \rev{\qty{82}{\percent}} reduction in walking time in all cases.
\end{abstract}
\input{intro}
\input{overview}
\input{Methods}
\input{Results}
\input{Discussion}
\input{conclusion}
\input{acknowledgments}

\newpage
\bibliographystyle{splncs04}
\bibliography{lit, bib_swissler}

\end{document}

%% file: intro.tex
\section{Introduction}

Although fixed-route public transit systems may effectively serve densely populated urban areas,
access to these networks is often limited to people who live and work in central residential and
business districts~\cite{taylor2013explaining,manville2018falling,chakrabarti2017can}. Demand-responsive
transit systems (DRTs), which adjust their routes and schedules according to passenger demand, have
been proposed as an alternative or supplement to fixed-route transit to increase transit accessibility,
particularly in communities with low population
density~\cite{mortazavi2024integrated,dytckov2022potential,coutinho2020impacts,calabro2025and,lakatos2020demand}
and lead to increased usage of other forms of public
transit~\cite{alonso2018potential,thao2023demand}. Despite these benefits, DRTs have a high
failure rate due to high costs and lack of reliability~\cite{currie2020most}.

Unlike fixed-route transit, DRTs must dynamically determine routes and schedules
to accommodate passenger demand. One approach to DRT routing is to use \rev{shortest-path}
algorithms inspired by Dijkstra's algorithm~\cite{dijkstra1959note}, with many optimized
implementations available~\cite{sommer2014shortest}. These algorithms are commonly used in
transit routing~\cite{zhan1997three,zhan1998shortest,zeng2009finding}, and have been adapted
to accommodate uncertain and stochastic networks~\cite{chen2013finding,chen2016finding},
and optimize for features in addition to path length~\cite{russ2021constrained}.
Index-based methods may be applied to speed up path search by pre-computing sub-paths within the
network~\cite{zhu2013shortest,ouyang2020efficient,qiu2022efficient,wang2019querying}.
Metaheuristic methods, which allow flexible optimization parameters but are not guaranteed to find
an optimal solution, have also been applied to the problem of routing on road networks, including
ant colony
optimization~\cite{cheng2023dynamic,dias2014inverted,singh2025optimization,goudarzi2018traffic},
particle swarm optimization~\cite{farahmand2024multi,marinakis2010hybrid},
tabu search~\cite{gmira2021tabu,ahmed2025optimal,venkatraman2021congestion},
and genetic
algorithms~\cite{huang2020flexible,kanoh2008hybrid,kanoh2007dynamic,dib2017combining,chakraborty2005multiobjective}.

We propose a novel distributed algorithm for demand-responsive transit, illustrated in
Figure~\ref{fig:introFig} and detailed in Section~\ref{sec:overview}, which aims to improve
efficiency, reliability, and quality-of-service over existing alternatives. Unlike shortest-path
approaches, this system uses the \rev{Robust Adaptive Pathfinding for Interconnected Devices (RAPID)}
algorithm \rev{described in Section~\ref{sec:rapid}}, a distributed algorithm inspired by the ability of the slime mold
\textit{Physarum polycephalum} to
identify efficient and robust paths through networks~\cite{nakagaki2000maze,tero2010rules} and make complex
decisions without any mechanism of centralized control~\cite{reid2015information,reid2023thoughts}.
The proposed system addresses many of the key limitations of current shortest-path based approaches.
\rev{
RAPID is not an optimization algorithm; it does not guarantee shortest paths determination, nor paths that are optimal in any other sense. Instead, RAPID makes the
best possible routing decision at each point in time
and space
using current locally available information,
adapting to changing conditions and requirements. The need for explicit planning is eliminated along with the computational burden and inflexibility of a highly planned system.
}

Studies of transit ridership indicate that many factors
in addition to transit time or travel distance have significant effects on ridership, including
wait time~\cite{he2022examining}, walking distance~\cite{taylor2013explaining}, number
of transfers, reliability, and predictability~\cite{chakrabarti2017can}. Therefore, a suitable routing
mechanism should incorporate all of these factors and allow tuning to balance relative priorities.
Even when considering constraints beyond distance, path-finding algorithms identify fixed paths
between a single source and destination.
Generating an adequate fully-planned route in advance, while accounting for traffic and changing
road conditions, remains a challenging problem~\cite{bast2016route} due to the scale and
unpredictability of real
road networks that cause these solutions to become sub-optimal or even infeasible in real time.
However, the DRT routing problem is even more complex
than determining an individual route, as multiple passengers must be routed simultaneously
using a limited number of vehicles, with the route for one passenger possibly affecting others. This
scenario leads to a high degree of uncertainty in which pre-planned routes are unlikely to remain
optimal or near-optimal, even in the short term. This problem is exacerbated when passengers are
allowed to request rides
in real-time rather than scheduling them long in advance, requiring the DRT routing system to
adapt quickly and frequently to unpredictable demand. As existing routing approaches require
centralized planning, the burden of computation and network communication for such a system becomes
infeasible for large transportation networks under high usage scenarios.

Rather than specifically computing individual paths, the RAPID algorithm used in the proposed system
efficiently creates
gradients of demand across the road network using only local network communication. This gradient is
maintained through repeated communications and thus reacts in real-time to changing demand
and network conditions. The gradient is used to inform local routing decisions by vehicles
without the need for advanced planning or coordination.
All components of the algorithm are
parameterized to allow \rev{tuning of system behavior}.
\rev{
We note that these parameters are not part of an objective function, but are directly used to influence
simple local decisions.
}

Using an agent-based simulation of both the proposed system and fixed-route transit, we find that
the RAPID-based system increases the number of passengers delivered in every simulation environment
relative to the fixed-route system. Additionally, \rev{walking time is reduced by
over \qty{82}{\percent}} in all cases.

The primary contributions of this work are: (1) the design of a novel, RAPID-based public transit
approach inspired by the intelligent behaviors of slime mold, (2) the development of a flexible
transit simulation that incorporates real world street and transit data, and (3) initial test
results and analysis that demonstrate the potential improvements of the proposed transit system.

%% file: overview.tex
\section{Narrative Overview of the System}

\label{sec:overview}

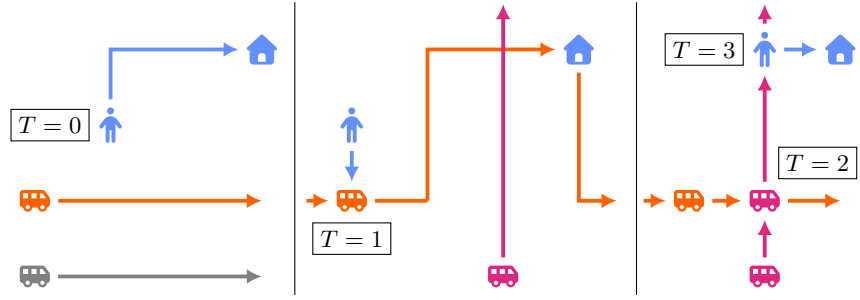
\begin{figure}[t]
  \centering
  \begin{subfigure}[b]{0.3\textwidth}
    \centering
    \begin{tikzpicture}
      \node (house) at (0,0) [house-blue];
      \node (person) at (-2,-1) [passenger-blue];
      \node (bus1) at (-3,-2) [van-shuttle-orange];
      \node (bus2) at (-3,-3) [van-shuttle-gray];

      \node at (-2.8,-1) [timecaption] {$T=0$};

      \draw[personarrow] (person) |- (house);
      \draw[orangearrow] (bus1) -- (0,-2);
      \draw[grayarrow] (bus2) -- (0,-3);
    \end{tikzpicture}
  \end{subfigure}
  \vrule
  \begin{subfigure}[b]{0.36\textwidth}
    \centering
    \begin{tikzpicture}
      \node (house) at (0,0) [house-blue];
      \node (person) at (-3,-1) [passenger-blue];
      \node (bus1-2) at (-3,-2) [van-shuttle-orange];
      \node (bus3) at (-1,-3) [van-shuttle-pink];

      \node at (-3,-2.5) [timecaption] {$T=1$};

      \draw[personarrow] (person) -- (bus1-2);
      \draw[orangearrow] (-3.6,-2) -- (bus1-2);
      \draw[orangeline] (bus1-2) -- (-2,-2);
      \draw[orangearrow] (-2,-2) |- (house);
      \draw[orangearrow] (house) |- (0.5,-2);
      \draw[pinkarrow] (bus3) -- (-1,0.6);
    \end{tikzpicture}
  \end{subfigure}
  \vrule
  \begin{subfigure}[b]{0.25\textwidth}
    \centering
    \begin{tikzpicture}
      \node (house) at (0,0) [house-blue];
      \node (bus1-2) at (-2,-2) [van-shuttle-orange];
      \node (bus3) at (-1,-3) [van-shuttle-pink];
      \node (bus3-2) at (-1,-2) [van-shuttle-pink];
      \node (person-2) at (-1,0) [passenger-blue];
      
      \node at (-0.3,-1.5) [timecaption] {$T=2$};
      \node at (-1.8,0) [timecaption] {$T=3$};

      \draw[orangearrow] (-2.6,-2) -- (bus1-2);
      \draw[orangearrow] (bus1-2) -- (bus3-2);
      \draw[orangearrow] (bus3-2) -- (0,-2);
      \draw[pinkarrow] (bus3) -- (bus3-2);
      \draw[pinkarrow] (bus3-2) -- (person-2);
      \draw[personarrow] (person-2) -- (house);
      \draw[pinkarrow] (person-2) -- (-1,0.6);
    \end{tikzpicture}
  \end{subfigure}
  \caption{\rev{A simplified example of Fluid Fleets behavior. Left: at time $T=0$, a passenger requests a ride and
  nearby vehicles bid to pick up the passenger. Middle: at time $T=1$, the closer vehicle picks up the
  passenger and updates its route. Right: at time $T=2$, the passenger transfers
  to another vehicle, leading to more efficient routes for both vehicles.}}
  \label{fig:introFig}
\end{figure}

Figure~\ref{fig:introFig} provides a \rev{simplified} example of how the system functions. Although the actual system runs on complex street networks pulled from real-world map data (upper left portion of the figure), here we present the system in an abstracted state with a focus on the interactions between the various agents. 

At $T=0$, a \rev{person initiates} a request to travel home. There are two \rev{vehicles that} the person could potentially use to get home; the arrows represent the current planned paths for the system. These vehicles then generate an individual bid for picking up the passenger, based on various factors such as proximity, route deviation, and \rev{capacity. They} compare bids and, in this case, the orange vehicle is selected to pick up the passenger.

After the vehicle is selected, the passenger and the vehicle both adjust their routes to meet at a designated pick-up location, with the passenger walking some short \rev{distance. At} $T=1$, the vehicle and the passenger meet and the passenger boards the vehicle. The route for the orange vehicle has become longer, increasing the travel time for the other passengers. 

\rev{During the trip, the} transfer mechanism (described in Section~\ref{sec:transfers}) \rev{evaluates} the efficiency of transferring the passenger \rev{to another vehicle} to determine \rev{whether to transfer}. In this case, \rev{a suitable transfer is identified} and the route of the orange vehicle is \rev{updated} to intercept the \rev{pink vehicle} at a location that results in an improved route for the orange vehicle while maintaining an efficient route for the passenger. 

At \rev{$T=2$}, the orange vehicle drops off the passenger, and a short time later the \rev{pink} vehicle picks them \rev{up. At} \rev{$T=3$}, the \rev{pink} vehicle drops off the passenger a short distance from their home, completing the trip.

%% file: methods.tex
\section{Methods}

We consider a network representation of a city in which nodes represent street intersections and
edges represent streets connecting intersections. The demand for transit is created by
passengers, who must be picked up at a particular node in the network and dropped off at a destination
node. \rev{The process for determining these values is \rev{described in} Section~\ref{sec:simulation} under Initialization}. \rev{Passengers} may walk to reach a pickup location for a transit
vehicle, or a destination from their vehicle's drop-off location. In the case of a fixed-route transit system, vehicles follow a fixed path that includes
certain nodes as pickup/drop-off locations. Passengers must walk to and from these locations to
reach their destination.

The RAPID algorithm approach does not provide fixed routes for the vehicles. Instead, each passenger
is assigned to a vehicle using a bidding mechanism which ensures that the assignment is optimized
according to a reward function described in Section~\ref{sec:bidding}. Each vehicle determines its route
according to the local gradients established by the RAPID algorithm, which responds in real-time to
changes in passenger demand and
network conditions, described in Section~\ref{sec:routing}. Passengers may also transfer between vehicles when
the transfer leads to a reduction in travel time, using the approach described in Section~\ref{sec:transfers}.

To evaluate the performance of the RAPID algorithm transit system and compare its effectiveness
against the fixed-route approach, we have implemented a simulation of these transit networks,
described in Section~\ref{sec:simulation}.

\rev{
\subsection{The RAPID algorithm}
\label{sec:rapid}

The Robust Adaptive Pathfinding for Interconnected Devices (RAPID) algorithm solves a problem we
call ``message-in-a-bottle'': transferring a message
to a designated recipient without the sender or the message knowing the destination or the route the
the message should take. The approach of RAPID is inspired by the behavior of the slime mold
\textit{P.\ polycephalum}, which is capable of reallocating its biomass toward nutrient sources
without the means to propagate information about nutrient locations~\cite{alim2018fluid}.
\textit{P.\ polycephalum} solves this problem by creating a pressure gradient leading to
nutrient-rich areas which directs the flow of its cytoplasm~\cite{gyllingberg2026self}.
Similarly,
the RAPID algorithm delivers objects on a network using only local communication along network edges
and without global knowledge of the location or existence of objects or destinations. Gradients
propagate through the network through peer-to-peer communication and originate at destination nodes,
resulting in a simulated pressure gradient similar to that in \textit{P.\ polycephalum}.

The RAPID algorithm implements three slime-mold inspired behaviors: (1) local information, namely
the status of a node as a destination, is represented through a numerical value indicating ``demand''
for an object, (2) communication with neighboring nodes influences the local demand value through
averaging, and (3) objects being routed move toward the neighbor with the highest demand value,
ultimately moving along the gradient toward the source of demand. The specific implementation of
the RAPID algorithm applied to Fluid Fleets is described in Section~\ref{sec:routing}. A more general
investigation of the capabilities of the RAPID algorithm is currently underway, including capabilities
not covered in this study such as dynamic networks and handling network congestion.
}

\subsection{Routing}
\label{sec:routing}

Vehicle routing in the proposed system uses the RAPID algorithm
\rev{described in Section~\ref{sec:rapid}.}
Each node $n$ has a demand value $d_{p,n}$ corresponding to
each passenger $p$ and a reference value $r_{p,n}$ which is used to update the demand. The reference
value is set to zero by default, and set to positive values at destinations, such as
pickup and drop-off locations. Demand is propagated along neighboring pairs $(m,n)$ of nodes according
to
\begin{align}
    d_{p,m} &= \zeta d_{p,n} + \eta r_{p,m} + \mu d_{p,m},\\
    d_{p,n} &= \zeta d_{p,m} + \eta r_{p,n} + \mu d_{p,n},
\end{align}
where $\zeta$, $\eta$, and $\mu$ are parameters controlling the
relative weights of the diffusion, relaxation, and inertia components
of the the demand propagation, respectively.
In order to create demand gradients leading to destinations, reference values of nodes where passengers
are assigned to be either picked up or dropped off are set to \num{10}. All other
reference values are set to \num{0}.

The cumulative demand value for each node in the vehicle's network is determined by taking the sum of the demand
values for each passenger assigned to the vehicle $i$, weighted by the cumulative amount of time the
passenger has spent waiting,
\begin{align}
    \hat{d}_{i,n} = \sum_{p\in P_i} (1 + \lambda T_p) d_{p,n},
\end{align}
where $P_i$ is the set of passengers assigned to vehicle $i$, $\lambda$ is a parameter which determines
the influence of the waiting time, and $T_p$ is the passenger waiting time. Routing decisions are
then locally determined at each node by the bus moving toward the adjacent node with the highest
demand value.

\subsection{Bidding}
\label{sec:bidding}

Passengers are assigned to vehicles using a bidding system.
Upon requesting a pickup, the requesting
passenger obtains a bid value for each candidate vehicle. The vehicle with the maximum bid value
is assigned to pick up the passenger. The bid value $B_{i,p}$ for each vehicle $i$ picking up
passenger $p$ is given by
\begin{align}
B_{i,p} = \alpha \overline{T}_p^\beta + \gamma \left( \frac{1}{\overline{D}_{i,p} + 1} \right)^\delta + \sigma \overline{N}_i^\epsilon,
\end{align}
where $\overline{T}_p$ is the normalized waiting time for passenger $p$, $\overline{D}_{i,p}$
is the normalized shortest-path distance between bus $i$ and passenger $p$, and $\overline{N}_i$
is the normalized number of passengers on bus $i$. Normalized values are computed as a proportion of
the maximum value for each quantity in the simulation. The parameters
$\alpha$, $\beta$, $\gamma$, $\delta$, $\sigma$, and $\epsilon$ allow tuning
of the reward function to promote desired behavior.

\rev{
Due to the focus of the current work
on the routing approach using the RAPID algorithm, the bidding process is currently
centralized for simplicity. Future work will explore distributed approaches to
vehicle assignment and evaluate their effectiveness compared with the centralized
scheme.
}

\subsection{Transfers}
\label{sec:transfers}

Transfers enable a passenger to switch from their current vehicle to another vehicle in cases where
the transfer will reduce their expected travel time. However, transfers require
both vehicles involved in the transfer to stop, potentially increasing travel time for other
passengers. This tradeoff is quantified using the bidding system described in
Section~\ref{sec:bidding} by measuring the change in the sum of the bid values for passengers
assigned to the vehicles. For passengers who have already been picked up, the location used for the
bid value is the destination location rather than the pickup location.

Transfers are only conducted when the following two conditions are met: first, the cumulative
improvement in bid values exceeds a threshold $\theta$, and second, the number of transfers for the
passenger has not exceeded a limit $\tau$.

\subsection{Simulation}
\label{sec:simulation}

We have implemented a simulation of two types of public transit network: the novel approach proposed
in this work, and fixed-route transit for the purpose of comparing the effectiveness of the two
approaches. The fixed-route transit simulation is
informed by data collected from the NJ Transit, the transit authority serving the state of New Jersey.

\subsubsection{Agents: Passengers and Vehicles}

The simulation used in this study includes two types of agents: passengers and vehicles. The
passengers in the simulation must track location, destination, number of previous transfers,
and timestamps of key events:
pickup request time, pickup times, dropoff times, walking time, and arrival time. Vehicles
must track their current position and maintain lists of currently assigned passengers,
with a distinction between those on board and those who have not yet been picked up.

\subsubsection{Simulation Environment}

\begin{figure}[t]
    \centering
    
        \includegraphics[width=0.7\textwidth,trim={90mm 58mm 75mm 65mm},clip]{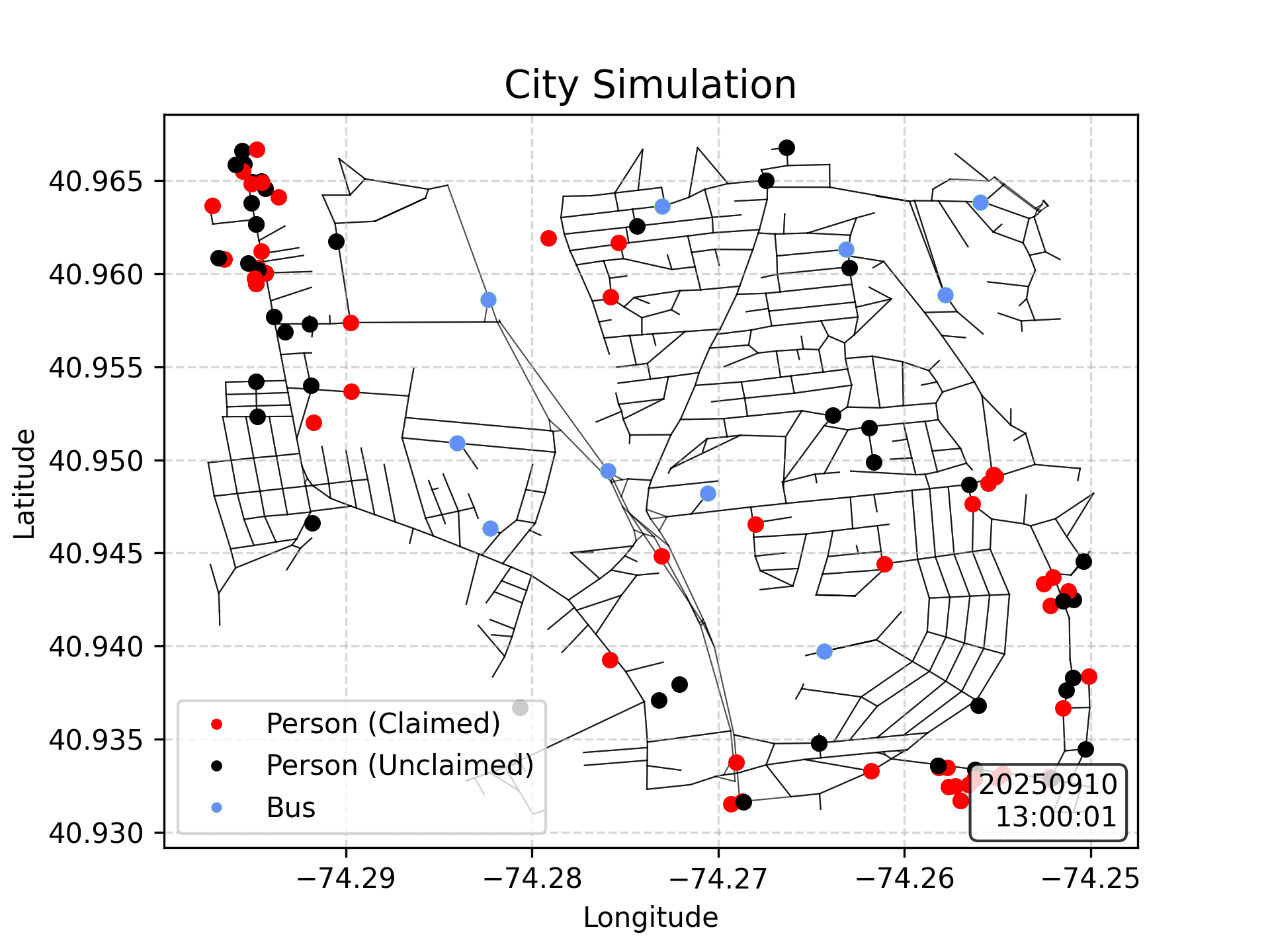}
    \caption{\rev{Screenshot of the Fluid Fleets simulation in the semi-rural environment. Black lines represent streets, passengers
    assigned to a vehicle are shown as red dots, passengers who have not yet been scheduled
    for pickup are shown as black dots, and buses are depicted as green dots.}}
    \label{fig:env_screenshot}
\end{figure}

The road networks used for the simulations are generated from geographical data taken from
OpenStreetMap. All roads are treated as straight lines for simplification. The environment is defined
by the center (latitude, longitude) coordinates in addition to a map radius to specify the size of
the environment. For simplicity, one way streets are not included in the simulated street network.

The environments used in our simulation were suburban Rutherford, NJ centered at (40.8263, -74.1071), semi-rural Wayne, NJ at (40.9489, -74.2736), and urban Newark, NJ at (40.7424, -74.1784), each with a 2 kilometer radius. \rev{An example simulation screenshot is} presented in Figure~\ref{fig:env_screenshot}. All three environments have existing public transportation infrastructure with NJ Transit, which enable us to compare existing routes with our RAPID-based routing method.

\subsubsection{Initialization}

The simulation is initialized with \num{100} people. Throughout
the simulation time, an additional person will be added at a fixed rate of 10 seconds, with a maximum of 300 people waiting to reach their destination at any given time.
The initial location for each person is randomly assigned using the following scheme: the map is
divided into a $10\times10$ grid. The grid cell for the person is chosen randomly, weighted by the
proportion of nodes within the cell. The initial location is then drawn from a uniform distribution
within the chosen cell.
People will then walk to the nearest street node along the walking map
obtained from OpenStreetMap at a constant speed randomly determined for each person, between 2 and 4 mph, inclusive.
\rev{
Person start and end locations have a minimum distance of
\qty{500}{\meter} from one another. Both locations are chosen within \qty{15}{\minute} walking
time to the nearest fixed-route stop for the individual passenger to ensure that
walking times remain reasonable for the fixed-route simulation.
This restriction reflects the fact that long walking times are a barrier to
transit usage~\cite{taylor2013explaining} for the fixed-route case, while placing the
same limitation on the Fluid Fleets simulation for a direct comparison. Future studies
will consider the effect of reduced walking times for many regions provided by Fluid Fleets,
which is expected to create a further advantage over fixed-route transit.
}

Vehicles in the RAPID-based simulation are initially assigned to random street nodes with equal
probability. Each vehicle moves at a constant speed of 30 mph along the street nodes.

\subsubsection{Simulation Parameters}

\rev{The simulation includes several tunable parameters that can be adjusted to control the behavior. The parameters and their values used for the simulation are listed in Table~\ref{tab:parameters}.}

\begin{table}[t]
    \centering
    \caption{Parameters used in the simulation.}
    \label{tab:parameters}
    \begin{tabular}{|c|c|c|}
         \hline
         Parameter & Description & Value\\
         \hline
         $\alpha$ & Bid wait weight & 0.05\\
         $\beta$ & Bid wait exponent & 0.94\\
         $\gamma$ & Bid distance weight & 10.00\\
         $\delta$ & Bid distance exponent & 2.54\\
         $\sigma$ & Bid passengers weight & 0.10\\
         $\epsilon$ & Bid passengers exponent & 0.96\\
         $\zeta$ & Demand diffusion weight & 0.40\\
         $\eta$ & Demand relaxation weight & 0.40\\
         $\mu$ & Demand inertia weight & 0.20\\
         $\lambda$ & Demand waiting scaling & 0.25\\
         $\omega$ & Genetic algorithm cost parameter & 0.75\\
         $\theta$ & Transfer threshold & 1.96\\
         $\tau$ & Maximum passenger transfers & 2\\
         \hline
    \end{tabular}
\end{table}

A genetic algorithm \cite{katoch2021review} was implemented to optimize
the parameters of the bidding function ($\alpha$, $\beta$, $\gamma$, $\delta$, $\sigma$, and $\epsilon$), wait time penalty ($\lambda$), and transfer threshold value ($\theta$) for the
simulation. The size of each
population was set to \num{1000}, with a keep ratio of \num{.1}, random ratio of \num{.1}, and mutation
rate of \num{.3}. The algorithm was run for a maximum of \num{1000} generations with early termination
after \num{100} generations without improvement. For exponents, the allowed range was -5.00 to 5.00. For weights, the allowed range was -10.00 to 10.00. The fitness function for the genetic algorithm is
\begin{align}
    F = \frac{1}{\omega t_m + (1-\omega) t_s},
\end{align}
where  $t_m$ is the mean passenger delivery time, $t_s$ is the standard deviation of delivery
times, and $\omega$ is \rev{a unitless parameter in the range $[0,1]$}.

\subsubsection{Fixed Routing Simulation}

In order to benchmark the efficacy of the proposed transit system, we compare the simulated results
against simulations of existing fixed-route transit networks in the same environments. The
fixed-route simulation consists of bus stops and routes determined from GTFS data downloaded through
the Mobility Database API. Simulated buses
travel continuously along their designated routes, with the routes and service times determined by
the available data. Passengers are initialized in the simulation using the same approach used for
the simulation of the RAPID-based system.

After initialization, passengers walk to the nearest bus stop that can service their destination within one transfer. A maximum of one bus transfer for passengers is allowed in the fixed routing simulation to reduce the time complexity of the transfer algorithm. Transfers are only conducted when the bus initially picking up the passenger does not have a stop at their drop off location, and there is an intersection point between the two routes. 

%% file: results.tex
\section{Results}

To evaluate the performance of the RAPID-based approach to public transit, we compare several
indicators of efficiency and quality-of-service with existing fixed-route
transit systems in various scenarios. For this study, we consider the number of cumulative number
of passengers delivered during the course of the simulation, which indicates the throughput of the
transit system under the simulated usage. As an indicator of quality-of-service and practical
usability, we also consider the mean walking distance required for the passengers in the
simulation.

\subsection{Comparison with Fixed-Route Transit in Various Scenarios}

\begin{figure}[t]
    \centering
    \includegraphics[width=\textwidth]{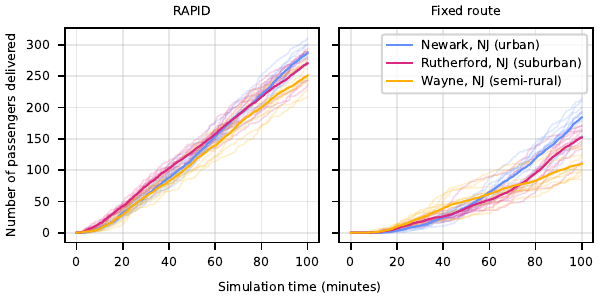}
    \caption{Passenger delivery results from RAPID-based routing (left) vs.\ traditional fixed routing (right) during a simulated \qty{100}{\minute} window. Thin, semi-transparent lines represent the results of individual simulations, with thicker opaque lines showing the average performance. \num{10} simulation runs were performed for each case.}
    \label{fig:routing}
\end{figure}

A comparison of the cumulative number of passengers delivered during \num{100} minutes of
simulation time for the transit systems in each community type is presented in
Figure~\ref{fig:routing}. All simulations used \num{10} vehicles with a capacity of \num{30}
passengers each. The RAPID-based system outperforms the fixed-route system in all cases,
notably attaining similar results in all three environments.
\rev{
These results are summarized
in Table~\ref{tab:routing}. Also included are differences in mean passenger walking time,
that Fluid Fleets reduces by over \qty{82}{\percent} in all cases.
}
This improvement highlights the benefits for transit access created by a flexible system in which
vehicles are able to pick up passengers at convenient locations rather than fixed stops.

\begin{table}[t]
    \centering
    \rev{
    \caption{
    \rev{
    Comparison of simulation results for fixed-route transit and Fluid Fleets. Number of
    deliveries over the \qty{100}{\minute} simulation are compared as well as the mean passenger
    walking time. Improvements (Imp.) are relative compared to the
    fixed-route case. Results are averaged over \num{10} simulation runs.
    }
    }
    \label{tab:routing}
    \vspace{2mm}
    \begin{tabular}{|c?{0.5mm}c|c|c?{0.5mm}c|c|c|}
         \hline
         & \multicolumn{3}{c?{0.5mm}}{Number of deliveries} & \multicolumn{3}{c|}{Mean walking time (min.) } \\
         \hline
         Location & Fixed-route & Fluid Fleets & Imp. & Fixed-route & Fluid Fleets & Imp. \\
         \hline
         Newark, NJ & \num{184.2} & \num{286.9} & \qty{55.8}{\percent} & \num{22.3} & \num{1.3} & \qty{94.0}{\percent} \\
         Rutherford, NJ & \num{152.3} & \num{270.6} & \qty{77.7}{\percent} & \num{17.7} & \num{1.3} & \qty{92.9}{\percent} \\
         Wayne, NJ & \num{110.3} & \num{251.3} & \qty{127.8}{\percent} & \num{18.4} & \num{2.9} & \qty{82.3}{\percent} \\
         \hline
    \end{tabular}
    }
\end{table}

\subsection{Effect of Fleet Vehicle Type}

\begin{figure}[!t]
    \centering
    \includegraphics[width=\textwidth]{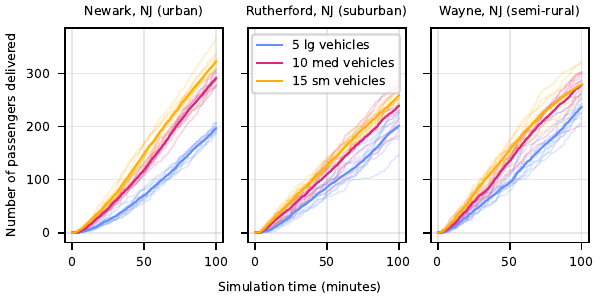}
    \caption{Passenger delivery results from various bus fleet configurations using the
    RAPID-based routing method during a simulated \num{100}-minute window.
    All simulated fleets have a total capacity of \num{300} passengers, with each case testing a
    different combination of number of vehicles and individual vehicle capacity.
    Thin, semi-transparent lines represent the results
    from individual simulations, with thicker opaque lines showing the average performance. \num{10} simulation runs were performed for each case.}
    \label{fig:location}
\end{figure}

The RAPID-based transit approach is designed to allow flexibility for fleet size and vehicle
capacity. Figure~\ref{fig:location} presents a comparison between various configurations. Vehicle
capacities are \num{60} passengers for the \num{5} vehicle case, \num{30} passengers for the \num{10}
vehicle case, and \num{20} passengers for the \num{15} vehicle case, resulting in a fixed total
capacity of \num{300} passengers.
We find similar trends in each scenario, with many smaller vehicles
resulting in faster delivery times than fewer larger vehicles.

%% file: discussion.tex
\section{Discussion}

We find that the RAPID-based approach outperforms existing fixed-route transit systems in the
simulated cases we tested. Additionally, we identify that the proposed system is particularly effective
when using fleets of \rev{smaller} vehicles, indicating that initial implementations may not require large
investments in expensive, bespoke vehicles. We also find that the RAPID-based approach results in
significantly reduced walking times in all tested areas, increasing accessibility of
public transit in these regions.
\rev{
These results show promise for Fluid Fleets to address the key issues of high cost and low reliability
that hinder adoption of DRTs. The RAPID algorithm underlying Fluid Fleets uses only a few interpretable
parameters, has low computational requirements, and adapts in real-time to changing conditions
as paths do not need to be planned or known ahead of time,
meeting the needs for a highly adaptable and reliable system. These advantages may
increase riders' perception of system reliability, a key
driver for adoption~\cite{chakrabarti2017can}. The ability to outperform fixed-route
transit using smaller vehicles with the same total capacity suggests that Fluid Fleets has the
potential to save money over underused fixed route transit systems, as we expect smaller vehicles to be
cheaper to purchase, fuel, and maintain.
}

Many simplifying assumptions were made for the simulation used in the study. Currently, the simulation
does not include traffic, which could significantly affect the overall behavior of both fixed-route
and RAPID-based transit. Other traffic conditions, such as speed limits, are also not directly considered,
with vehicles moving at constant speed in all cases. Including one-way streets may allow
improved routes for the RAPID-based system by allowing more routing options in certain scenarios,
while other limitations such the limited ability of certain streets to accommodate larger vehicles
may limit possible routes. Passenger locations and destinations are
currently randomly determined, while real transit usage is likely to follow trends such as movement
to and from urban centers during commuting times.

\rev{
This study uses fixed-route transit as a performance benchmark due the the ready availability of data in the areas of interest. However, comparisons with other DRT approaches and
other methods of routing should be established in future work, especially optimization-based
algorithms and those that require centralized planning. In addition to system performance, factors
such as computational requirements and the assumptions required for successful system performance
should also be considered.
}

Parameters selected by the genetic algorithm were
determined only from the Newark, NJ simulation \rev{environment,
so} it is possible that parameters determined from each location separately could achieve
improved results. The implementation of the genetic algorithm could also be improved. A
more complex fitness function may allow additional metrics to be optimized. The current range of
parameter values may also be inappropriate as the bid distance weight is set to the maximum limit,
indicating that either a higher value or the inclusion of a regularization term may result in an
improved parameterization. Ongoing work is focused on improving the performance of the simulation,
which will allow larger populations for the genetic algorithm for improved fitting results without
requiring excessive computational resources.

Estimated walking times for fixed-route transit simulations are most likely large because the simulation
requires all passengers use public transit, while real potential passengers often avoid public transit
entirely if it is too inconvenient. A possible improvement to this approach would be to incorporate
the decision of whether or not to take transit for each passenger, with factors such as required walking
and travel time influencing the decision, to directly measure the potential for the system to increase
transit usage.

Additionally,
in order to keep the simulation size reasonable for initial trials, only a subset of the existing
transit networks are considered in each case. A full, city-wide simulation would require significantly
more computational work but may provide new insights.
Finally, a key issue for DRTs is cost, which is not evaluated in the present study. We \rev{will}
include cost \rev{and revenue}
metrics in the future to establish the most reasonable comparisons between the RAPID-based transit
approach and others, with a focus on identifying cases where the proposed system can introduce
cost-saving. \rev{This information could be used to estimate the effects
of various system parameters on system cost, allowing transit authorities to identify the most
cost-effective settings for their environment.}

Due to the initial promising results from the smaller-scale simulations found in this study, we
plan to incorporate many of these missing components in the future to further validate the approach.
Additionally, longer simulations that take place
over a full day could demonstrate longer-term effects or patterns. An inherent advantage of
the RAPID-based approach is the ability to expand outside the existing network without the need for
planning existing routes or building additional infrastructure, but these capabilities have yet to be
tested. Furthermore, a common use of DRTs is to extend, rather than replace, existing transit systems,
so future simulations could consider the effect of adding the RAPID-based approach in addition to
fixed-route transit, with a particular focus on the relative benefits and efficiency of allocating
further resources toward additional fixed-route capacity and coverage compared with adding the proposed
system.

\rev{
In addition to future simulation work, we have begun working with transit authorities to plan
pilot studies of the Fluid Fleets. As Fluid Fleets does not require specialized vehicles or hardware
beyond what is already used in current DRT implementations, the primary concern to be addressed by
pilot studies is the effectiveness of the algorithm behavior under real-world limitations. A significant
practical challenge is the need to implement the algorithm software within the existing software
used by the transit authorities. Our approach to overcoming this challenge is to work closely
with transit authorities to implement Fluid Fleets within their system.
}

%% file: conclusion.tex
\section{Conclusion}

We present a novel demand-responsive transit approach inspired by the \rev{decentralized} decision-making
abilities of slime mold.
Our initial simulation results show that this system outperforms existing
fixed-route transit networks in all cases we tested, both in terms of passengers delivered in a fixed
time window and the required walking time for the passengers. We believe that the RAPID approach to
public transit is a promising new direction that will enable increased efficiency, lower cost, and
improved quality of service, ultimately leading to sustainable transit networks and improving
public transit adoption. Additionally, we believe that the flexibility and low barrier to implementation
of this system makes it particularly well-suited for rural and suburban environments which are not
well-served by fixed-route public transit.

%% file: acknowledgments.tex
\begin{credits}
\subsubsection{\ackname} 
This work was supported by an NJIT Faculty Seed Grant as well as by the National Science Foundation Award \#2452560.

\subsubsection{\discintname}
Authors declare no competing interests.
\end{credits}

%% file: lit.bib
@article{alonso2018potential,
  title={The potential of demand-responsive transport as a complement to public transport: An assessment framework and an empirical evaluation},
  author={Alonso-Gonz{\'a}lez, Mar{\'\i}a J and Liu, Theo and Cats, Oded and Van Oort, Niels and Hoogendoorn, Serge},
  journal={Transportation Research Record},
  volume={2672},
  number={8},
  pages={879--889},
  year={2018},
  publisher={SAGE Publications Sage CA: Los Angeles, CA}
}

@article{mortazavi2024integrated,
  title={Integrated demand responsive transport in low-demand areas: a case study of {Canberra}, {Australia}},
  author={Mortazavi, Amir and Ghasri, Milad and Ray, Tapabrata},
  journal={Transportation Research Part D: Transport and Environment},
  volume={127},
  pages={104036},
  year={2024},
  publisher={Elsevier}
}

@article{dytckov2022potential,
  title={Potential benefits of demand responsive transport in rural areas: A simulation study in {Lolland}, {Denmark}},
  author={Dytckov, Sergei and Persson, Jan A and Lorig, Fabian and Davidsson, Paul},
  journal={Sustainability},
  volume={14},
  number={6},
  pages={3252},
  year={2022},
  publisher={MDPI}
}

@article{thao2023demand,
  title={Demand responsive transport: New insights from peri-urban experiences},
  author={Thao, Vu Thi and Imhof, Sebastian and von Arx, Widar},
  journal={Travel Behaviour and Society},
  volume={31},
  pages={141--150},
  year={2023},
  publisher={Elsevier}
}

@article{coutinho2020impacts,
  title={Impacts of replacing a fixed public transport line by a demand responsive transport system: Case study of a rural area in {Amsterdam}},
  author={Coutinho, Felipe Mariz and van Oort, Niels and Christoforou, Zoi and Alonso-Gonz{\'a}lez, Mar{\'\i}a J and Cats, Oded and Hoogendoorn, Serge},
  journal={Research in Transportation Economics},
  volume={83},
  pages={100910},
  year={2020},
  publisher={Elsevier}
}

@article{lakatos2020demand,
  title={Demand responsive transport service of ‘dead-end villages’ in interurban traffic},
  author={Lakatos, Andr{\'a}s and T{\'o}th, J{\'a}nos and M{\'a}ndoki, P{\'e}ter},
  journal={Sustainability},
  volume={12},
  number={9},
  pages={3820},
  year={2020},
  publisher={MDPI}
}

@article{calabro2025and,
  title={Where and When does Demand-Responsive Transport work best? A Parametric Analysis using Agent-based Modelling},
  author={Calabr{\`o}, Giovanni and Le Pira, Michela and Inturri, Giuseppe},
  journal={EURO Journal on Transportation and Logistics},
  pages={100173},
  year={2025},
  publisher={Elsevier}
}

@article{he2022examining,
  title={Examining the factors influencing microtransit users’ next ride decisions using Bayesian networks},
  author={He, Jiajing and Ma, Tai-Yu},
  journal={European Transport Research Review},
  volume={14},
  number={1},
  pages={47},
  year={2022},
  publisher={Springer}
}

@article{currie2020most,
  title={Why most DRT/Micro-Transits fail--What the survivors tell us about progress},
  author={Currie, Graham and Fournier, Nicholas},
  journal={Research in Transportation Economics},
  volume={83},
  pages={100895},
  year={2020},
  publisher={Elsevier}
}

@article{taylor2013explaining,
  title={Explaining transit ridership: What has the evidence shown?},
  author={Taylor, Brian D and Fink, Camille NY},
  journal={Transportation Letters},
  volume={5},
  number={1},
  pages={15--26},
  year={2013},
  publisher={Taylor \& Francis}
}

@article{manville2018falling,
  title={Falling transit ridership: California and Southern California},
  author={Manville, Michael and Taylor, Brian D and Blumenberg, Evelyn},
  year={2018}
}

@article{chakrabarti2017can,
  title={How can public transit get people out of their cars? An analysis of transit mode choice for commute trips in Los Angeles},
  author={Chakrabarti, Sandip},
  journal={Transport Policy},
  volume={54},
  pages={80--89},
  year={2017},
  publisher={Elsevier}
}

@article{nakagaki2000maze,
  title={Maze-solving by an amoeboid organism},
  author={Nakagaki, Toshiyuki and Yamada, Hiroyasu and T{\'o}th, {\'A}gota},
  journal={Nature},
  volume={407},
  number={6803},
  pages={470--470},
  year={2000},
  publisher={Nature Publishing Group UK London}
}

@article{tero2010rules,
  title={Rules for biologically inspired adaptive network design},
  author={Tero, Atsushi and Takagi, Seiji and Saigusa, Tetsu and Ito, Kentaro and Bebber, Dan P and Fricker, Mark D and Yumiki, Kenji and Kobayashi, Ryo and Nakagaki, Toshiyuki},
  journal={Science},
  volume={327},
  number={5964},
  pages={439--442},
  year={2010},
  publisher={American Association for the Advancement of Science}
}

@article{reid2015information,
  title={Information integration and multiattribute decision making in non-neuronal organisms},
  author={Reid, Chris R and Garnier, Simon and Beekman, Madeleine and Latty, Tanya},
  journal={Animal Behaviour},
  volume={100},
  pages={44--50},
  year={2015},
  publisher={Elsevier}
}

@article{alim2018fluid,
  title={Fluid flows shaping organism morphology},
  author={Alim, Karen},
  journal={Philosophical Transactions of the Royal Society B: Biological Sciences},
  volume={373},
  number={1747},
  pages={20170112},
  year={2018},
  publisher={The Royal Society}
}

@article{reid2023thoughts,
  title={Thoughts from the forest floor: a review of cognition in the slime mould Physarum polycephalum},
  author={Reid, Chris R},
  journal={Animal cognition},
  volume={26},
  number={6},
  pages={1783--1797},
  year={2023},
  publisher={Springer}
}

@article{dijkstra1959note,
  title={A note on two problems in connexion with graphs},
  author={Dijkstra, EW},
  journal={Numerische Mathematik},
  volume={1},
  number={1},
  pages={269--271},
  year={1959}
}

@article{zhan1998shortest,
  title={Shortest path algorithms: an evaluation using real road networks},
  author={Zhan, F Benjamin and Noon, Charles E},
  journal={Transportation science},
  volume={32},
  number={1},
  pages={65--73},
  year={1998},
  publisher={INFORMS}
}

@article{zhan1997three,
  title={Three fastest shortest path algorithms on real road networks: Data structures and procedures},
  author={Zhan, F Benjamin and others},
  journal={Journal of geographic information and decision analysis},
  volume={1},
  number={1},
  pages={69--82},
  year={1997}
}

@article{zeng2009finding,
  title={Finding shortest paths on real road networks: the case for A},
  author={Zeng, Wei and Church, Richard L},
  journal={International journal of geographical information science},
  volume={23},
  number={4},
  pages={531--543},
  year={2009},
  publisher={Taylor \& Francis}
}

@inproceedings{zhu2013shortest,
  title={Shortest path and distance queries on road networks: towards bridging theory and practice},
  author={Zhu, Andy Diwen and Ma, Hui and Xiao, Xiaokui and Luo, Siqiang and Tang, Youze and Zhou, Shuigeng},
  booktitle={Proceedings of the 2013 ACM SIGMOD International Conference on Management of Data},
  pages={857--868},
  year={2013}
}

@incollection{bast2016route,
  title={Route planning in transportation networks},
  author={Bast, Hannah and Delling, Daniel and Goldberg, Andrew and M{\"u}ller-Hannemann, Matthias and Pajor, Thomas and Sanders, Peter and Wagner, Dorothea and Werneck, Renato F},
  booktitle={Algorithm engineering: Selected results and surveys},
  pages={19--80},
  year={2016},
  publisher={Springer}
}

@article{sommer2014shortest,
  title={Shortest-path queries in static networks},
  author={Sommer, Christian},
  journal={ACM Computing Surveys (CSUR)},
  volume={46},
  number={4},
  pages={1--31},
  year={2014},
  publisher={ACM New York, NY, USA}
}

@article{chen2013finding,
  title={Finding reliable shortest paths in road networks under uncertainty},
  author={Chen, Bi Yu and Lam, William HK and Sumalee, Agachai and Li, Qingquan and Shao, Hu and Fang, Zhixiang},
  journal={Networks and spatial economics},
  volume={13},
  number={2},
  pages={123--148},
  year={2013},
  publisher={Springer}
}

@article{ouyang2020efficient,
  title={Efficient shortest path index maintenance on dynamic road networks with theoretical guarantees},
  author={Ouyang, Dian and Yuan, Long and Qin, Lu and Chang, Lijun and Zhang, Ying and Lin, Xuemin},
  journal={Proceedings of the VLDB Endowment},
  volume={13},
  number={5},
  pages={602--615},
  year={2020},
  publisher={VLDB Endowment}
}

@article{russ2021constrained,
  title={The constrained reliable shortest path problem in stochastic time-dependent networks},
  author={Ru{\ss}, Matthias and Gust, Gunther and Neumann, Dirk},
  journal={Operations Research},
  volume={69},
  number={3},
  pages={709--726},
  year={2021},
  publisher={INFORMS}
}

@article{chen2016finding,
  title={Finding the k reliable shortest paths under travel time uncertainty},
  author={Chen, Bi Yu and Li, Qingquan and Lam, William HK},
  journal={Transportation Research Part B: Methodological},
  volume={94},
  pages={189--203},
  year={2016},
  publisher={Elsevier}
}

@article{qiu2022efficient,
  title={Efficient shortest path counting on large road networks},
  author={Qiu, Yu-Xuan and Wen, Dong and Qin, Lu and Li, Wentao and Li, Rong-Hua and others},
  journal={Proceedings of the VLDB Endowment},
  year={2022},
  publisher={Association for Computing Machinery (ACM)}
}

@article{wang2019querying,
  title={Querying Shortest Paths on Time Dependent Road Networks.},
  author={Wang, Yong and Li, Guoliang and Tang, Nan},
  journal={Proc. VLDB Endow.},
  volume={12},
  number={11},
  pages={1249--1261},
  year={2019}
}

@article{huang2020flexible,
  title={Flexible route optimization for demand-responsive public transit service},
  author={Huang, Ailing and Dou, Ziqi and Qi, Liuzi and Wang, Lewen},
  journal={Journal of Transportation Engineering, Part A: Systems},
  volume={146},
  number={12},
  pages={04020132},
  year={2020},
  publisher={American Society of Civil Engineers}
}

@article{cheng2023dynamic,
  title={Dynamic path optimization based on improved ant colony algorithm},
  author={Cheng, Juan},
  journal={Journal of Advanced Transportation},
  volume={2023},
  number={1},
  pages={7651100},
  year={2023},
  publisher={Wiley Online Library}
}

@article{dias2014inverted,
  title={An inverted ant colony optimization approach to traffic},
  author={Dias, Jos{\'e} Capela and Machado, Penousal and Silva, Daniel Castro and Abreu, Pedro Henriques},
  journal={Engineering Applications of Artificial Intelligence},
  volume={36},
  pages={122--133},
  year={2014},
  publisher={Elsevier}
}

@article{singh2025optimization,
  title={Optimization of Path for Road Network With Modified Ant Colony Optimization (MACO)},
  author={Singh, Raushan Kumar and Kumar, Mukesh},
  journal={Concurrency and Computation: Practice and Experience},
  volume={37},
  number={3},
  pages={e8375},
  year={2025},
  publisher={Wiley Online Library}
}

@article{goudarzi2018traffic,
  title={Traffic-aware VANET routing for city environments—A protocol based on ant colony optimization},
  author={Goudarzi, Forough and Asgari, Hamid and Al-Raweshidy, Hamed S},
  journal={IEEE Systems Journal},
  volume={13},
  number={1},
  pages={571--581},
  year={2018},
  publisher={IEEE}
}

@incollection{farahmand2024multi,
  title={Multi-modal routing in urban transportation network using multi-objective quantum particle swarm optimization},
  author={Farahmand-Tabar, Salar and Afrasyabi, Parastoo},
  booktitle={Applied Multi-objective Optimization},
  pages={133--154},
  year={2024},
  publisher={Springer}
}

@article{marinakis2010hybrid,
  title={A hybrid particle swarm optimization algorithm for the vehicle routing problem},
  author={Marinakis, Yannis and Marinaki, Magdalene and Dounias, Georgios},
  journal={Engineering Applications of Artificial Intelligence},
  volume={23},
  number={4},
  pages={463--472},
  year={2010},
  publisher={Elsevier}
}

@article{gmira2021tabu,
  title={Tabu search for the time-dependent vehicle routing problem with time windows on a road network},
  author={Gmira, Maha and Gendreau, Michel and Lodi, Andrea and Potvin, Jean-Yves},
  journal={European Journal of Operational Research},
  volume={288},
  number={1},
  pages={129--140},
  year={2021},
  publisher={Elsevier}
}

@article{ahmed2025optimal,
  title={Optimal path recommendation in dynamic traffic networks using the hybrid Tabu-A* algorithm},
  author={Ahmed, Gamil and Sheltami, Tarek and Yasar, Ansar},
  journal={Transportation Research Part E: Logistics and Transportation Review},
  volume={204},
  pages={104414},
  year={2025},
  publisher={Elsevier}
}

@article{venkatraman2021congestion,
  title={A congestion-aware Tabu search heuristic to solve the shared autonomous vehicle routing problem},
  author={Venkatraman, Prashanth and Levin, Michael W},
  journal={Journal of Intelligent Transportation Systems},
  volume={25},
  number={4},
  pages={343--355},
  year={2021},
  publisher={Taylor \& Francis}
}

@inproceedings{kanoh2008hybrid,
  title={Hybrid genetic algorithm for dynamic multi-objective route planning with predicted traffic in a real-world road network},
  author={Kanoh, Hitoshi and Hara, Kenta},
  booktitle={Proceedings of the 10th annual conference on Genetic and evolutionary computation},
  pages={657--664},
  year={2008}
}

@article{dib2017combining,
  title={Combining VNS with genetic algorithm to solve the one-to-one routing issue in road networks},
  author={Dib, Omar and Manier, Marie-Ange and Moalic, Laurent and Caminada, Alexandre},
  journal={Computers \& Operations Research},
  volume={78},
  pages={420--430},
  year={2017},
  publisher={Elsevier}
}

@article{kanoh2007dynamic,
  title={Dynamic route planning for car navigation systems using virus genetic algorithms},
  author={Kanoh, Hitoshi},
  journal={International Journal of Knowledge-based and Intelligent Engineering Systems},
  volume={11},
  number={1},
  pages={65--78},
  year={2007},
  publisher={SAGE Publications Sage UK: London, England}
}

@inproceedings{chakraborty2005multiobjective,
  title={Multiobjective route selection for car navigation system using genetic algorithm},
  author={Chakraborty, Basabi and Maeda, Takeaki and Chakraborty, Goutam},
  booktitle={Proceedings of the 2005 IEEE Midnight-Summer Workshop on Soft Computing in Industrial Applications, 2005. SMCia/05.},
  pages={190--195},
  year={2005},
  organization={IEEE}
}

@article{katoch2021review,
  title={A review on genetic algorithm: past, present, and future},
  author={Katoch, Sourabh and Chauhan, Sumit Singh and Kumar, Vijay},
  journal={Multimedia tools and applications},
  volume={80},
  number={5},
  pages={8091--8126},
  year={2021},
  publisher={Springer}
}

@article{gyllingberg2026self,
  title={Self-organizing physical and biochemical interactions explain diverse behaviours in Physarum polycephalum},
  author={Gyllingberg, Linn{\'e}a and Haque, Abid and Ray, Subash K and Weber, Gregory and Graham, Jason M and Garnier, Simon},
  journal={bioRxiv},
  pages={2026--05},
  year={2026},
  publisher={Cold Spring Harbor Laboratory}
}
